# Analytic Density of States in the Abrikosov-Gorkov Theory

R. V. A. Srivastava and W. Teizer

*Department of Physics, Texas A&M University, College Station, TX 77843-4242, USA*

## ABSTRACT

Since the early 1960s, Abrikosov-Gorkov theory has been used to describe superconductors with paramagnetic impurities. Interestingly, the density of states resulting from the theoretical framework has to date only been known approximately, as a numeric solution of a complex polynomial. Here we introduce an exact analytic solution for the density of states of a superconductor with paramagnetic impurities. The solution is valid in the whole regime of Abrikosov-Gorkov theory; both where there is an energy gap and gapless. While of fundamental interest, we argue that this solution also has computational benefits in the evaluation of integrals for tunneling conductances and allows for an analytic description of materials with densities of states that are modeled from the basic Abrikosov-Gorkov density of states.

1. INTRODUCTION

The density of states (DOS) of a clean, conventional superconductor is customarily described by the Bardeen-Cooper-Schrieffer[1] (BCS) theory, based on a model of paired electrons. Bogoliubov et al.[2] and Valatin[3] independently found a canonical transformation of the Hamiltonian, thus simplifying the mathematics of the BCS theory. Using this, Abrikosov et al.[4] decoupled the appropriate Green's functions equations of motion to use and extend BCS theory for external potentials and gauge invariance, which allows a description of strong magnetic field and impurities in the system.

Abrikosov and Gorkov[4,5] (AG) used the BCS model interaction and described the impurities as point scattering sources with short-ranged effects. These impurities can change the electromagnetic properties of superconductors by removing the anisotropy, changing the electronic density of states and even altering the effective interaction.[6] The exchange interaction between the impurity spins and conduction electrons give an indirect exchange coupling between the impurity spins which can lead to magnetic ordering at low temperature.[7,8]

The AG DOS typically is derived by numerically evaluating the solutions of a complex, fourth order polynomial:

$$u^4 - 2u^3\xi + u^2(\xi^2 + \alpha^2 - 1) + 2u\xi - \xi^2 = 0, \qquad (1)$$

where $\xi$ represents the energy difference to the Fermi energy and $\alpha$ is a parameter. One of the roots u (selected according to the boundary conditions) is then used to numerically construct the DOS $\rho$ according to

$$\rho = N(u) = \mathrm{Re}\left[\frac{u}{(u^2-1)^{1/2}}\right]. \qquad (2)$$

This approach is quite inconvenient and requires significant computational power in projects where numerical evaluation of folding integrals with Fermi functions is required due to finite temperature, e.g. in the case of evaluating tunneling conductances. To our knowledge, no analytic expression of the AG DOS has been described to date.

## 2. RESULTS

AG theory describes superconductors with paramagnetic impurities and recent work[9] is aimed at a more specific microscopic understanding of the superconductor near these impurities. In spite of these recent efforts, the availability of an exact DOS in AG theory is important for ongoing work in several systems. We have constructed this analytical solution of the AG DOS from an analytic form for u which we found:

$$u = \frac{1}{6}\left(3\xi + \sqrt{3}\sqrt{\left(A + \frac{B}{E} + E\right)} - \sqrt{3}\sqrt{2A - \frac{B}{E} - E - \frac{6\sqrt{3}(1+\alpha^2)\xi}{\sqrt{\left(A + \frac{B}{E} + E\right)}}}\right), \quad (3)$$

where the values of A,B,C,D and E are given as:

$$\begin{aligned}
A &= 2 - 2\alpha^2 + \xi^2, \\
B &= \left(-1 + \alpha^2 + \xi^2\right)^2, \\
C &= -1 + \alpha^6 + 3\xi^2 - 3\xi^4 + \xi^6 + 3\alpha^4\left(-1 + \xi^2\right) + 3\alpha^2\left(1 + 16\xi^2 + \xi^4\right), \\
D &= 6\sqrt{3}\sqrt{\alpha^2\xi^2\left(\alpha^6 + 3\alpha^4\left(-1 + \xi^2\right) + \left(-1 + \xi^2\right)^3 + 3\alpha^2\left(1 + 7\xi^2 + \xi^4\right)\right)}, \\
E &= (C + D)^{\frac{1}{3}}.
\end{aligned} \quad (4)$$

This expression can readily be introduced into Eq. 1 leading to an identical cancellation and thus constitutes a solution of the equation. Fig. 1 shows plots of ρ=N(u) (using u from Eq. 3) versus ξ which are reminiscent of numerical solutions used for decades;[10] however the curves shown are exact functional solutions.

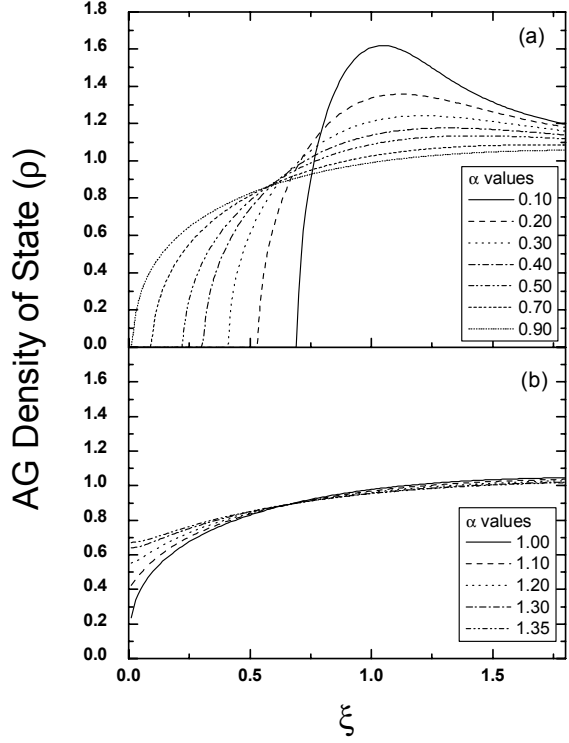

**FIG. 1. AG density of states vs. normalized energy for different values of α, (a) α<1 and (b) α≥1 ("gapless").**

The general analytic form, with free parameters $\xi$ and $\alpha$ is a significant step forward from a numeric evaluation. Furthermore, in some systems a functional combination of several AG DOSs is needed. Again this is significantly aided by the availability of an analytic form.

## 3. CONCLUSIONS

We have found an analytical solution of AG theory. This allows the construction of more complex theoretical DOSs, e.g. the combinations of several AG DOS. In addition, the analytical solution does not require a verification of the solutions independence of the numeric process.

ACKNOWLEDGEMENTS

We thank the Robert A. Welch Foundation (A-1585) for financial support and Chia-Ren Hu and Robert C. Dynes for valuable discussions and comments.

REFERENCES

[1] J. Bardeen, L. N. Cooper, and J. R. Schrieffer, Phys. Rev. **108**, 1175 (1957).
[2] N. N. Bogoliubov, V. V. Tolmachev, D. V. Shrikov, Fortschr. Physik **6**, 605 (1958).
[3] J. G. Valatin, Nuovo Cimento **7**, 405 (1958).
[4] A. A. Abrikosov, L. P. Gor'kov, and I. E. Dzyaloshinki, *Methods of Quantum Field Theory in Statistical Physics* (Prentice-Hall, inc., Englewood Cliffs, New Jersey, 1963), chap. 7.
[5] A. A. Abrikosov and L. P. Gor'kov, Zh. Eksp. Teor. Fiz. **39**, 1781 (1960); **39**, 866 (1961) [Soviet Phys.—JETP **12**, 1243 (1961)].
[6] C. Caroli, P. G. De Gennes, and J. Matricon, J. Phys. Radium **23**, 707 (1962).
[7] P. G. De Gennes and G. Sarma, J. Appl. Phys. **34**, 1380 (1963).
[8] L. P. Gorkov and A. I. Rusinov, J. Exptl. Theoret. Phys. U.S.S.R. **46**, 1363 (1964).
[9] E.W. Hudson *et al*, *Science* **285**, 88 (1999).
[10] A. Baratoff, PhD. thesis, Cornell University, 1964, Pg. 77.